\documentclass[reprint, aps, pra, twoside, floatfix, nofootinbib]{revtex4-1}
\usepackage{amsmath, amssymb}
\usepackage{graphicx}
\usepackage[usenames, dvipsnames]{color}
\usepackage{hyperref}
\hypersetup{colorlinks=true, linkcolor = blue, citecolor = blue, urlcolor = blue}
\newcommand{\opB}{\beta}
\newcommand{\opO}{\hat{\mathcal{O}}}
\newcommand{\opH}{\hat{\mathcal{H}}}

\newcommand{\sigmax}{\sigma_\mathrm{x}}
\newcommand{\sigmay}{\sigma_\mathrm{y}}
\newcommand{\sigmaz}{\sigma_\mathrm{z}}

\newcommand{\tE}{\mathcal{E}}
\newcommand{\tR}{\mathcal{R}}
\newcommand{\omegaa}{\omega_\mathrm{a}}
\newcommand{\omegac}{\omega_\mathrm{c}}
\newcommand{\tomegac}{\widetilde{\omega}_\mathrm{c}}
\newcommand{\Deltaa}{\Delta_\mathrm{a}}
\newcommand{\tDeltaa}{\widetilde{\Delta}_\mathrm{a}}
\newcommand{\dd}{\mathrm{d}}
\newcommand{\ee}{\mathrm{e}}
\newcommand{\ii}{\mathrm{i}}
\newcommand{\vacket}{|\varnothing\rangle}
\newcommand{\vacketw}{|\varnothing_\mathrm{w}\rangle}
\newcommand{\vacketc}{|\varnothing_\mathrm{c}\rangle}
\newcommand{\vacketjc}{|\varnothing_\mathrm{JC}\rangle}
\newcommand{\vacbra}{\langle\varnothing|}
\everymath{\displaystyle}
\begin{document}
\title{Few-photon Fock-state wave packet interacting with a cavity-atom system in a waveguide: Exact quantum state dynamics}
%
\author{E.~V. Stolyarov}
%
\email{eugenestolyarov@gmail.com}
\affiliation{Institute of Physics, National Academy of Sciences of Ukraine, Prospekt Nauky 46, 03028 Kyiv, Ukraine}
\begin{abstract}
	In the paper, we employ a wavefunction approach to investigate the evolution of a two-photon wave packet propagating in a one-dimensional waveguide coupled to the Jaynes-Cummings (JC) system.
	We derive and solve, both analytically and numerically, a set of equations of motion governing the quantum state of the system.
	That allows us to provide real-time analysis of the evolution of the wave packet two-photon joint spectrum (2PJS) and the excitation dynamics of the JC system in the course of its interaction with the two-photon pulse.
	We demonstrate that the 2PJS and the spectrum of the wave packet scattered from the JC system experience transformation for nonzero atom-cavity couplings.
	Moreover, using Schmidt decomposition, we show that the scattered photons feature frequency entanglement contrary to the incident ones which are not entangled.
\end{abstract}
%
%
\setcounter{page}{1}
\maketitle
\section{Introduction} \label{sec:intro}
Propagating photons can act as mobile qubits carrying quantum information between stationary qubits, where it can be stored or subjected to quantum logical operations \cite{north}. Remote stationary qubits are interconnected via waveguides (either microwave or optical ones) ensuring almost lossless transport of photons within quantum information processing (QIP) devices. Waveguides confine light in a transverse direction forming an effective one-dimensional (1D) continuum for propagating photons. Due to the Purcell effect, this modified electromagnetic environment gives rise to enhanced light-matter interaction. Strong light-matter coupling enables fast and accurate quantum-state transfer between photons and stationary qubits, which is essential for the realization of high-fidelity single- and multi-qubit gates.
 
Remarkable progress in creation and manipulation of waveguide quantum electrodynamics (wQED) systems was achieved in the last couple of decades.
In addition, several designs of single- and few-photon sources operating in the microwave \cite{hou,hof,pfdiaz,pfaff} and optical \cite{claud,snij,hansch} domains were demonstrated experimentally. That makes wQED a highly promising and versatile hardware platform for implementation of QIP systems. Ongoing technological advances in this field galvanize extensive theoretical studies of wQED systems. Several wQED-based architectures for quantum computation were proposed recently \cite{zheng2013, paulisch, pichler}.
Diverse theoretical techniques were engaged for studying a photon propagation in various wQED setups. 
Below we present a brief overview of some of these approaches.
 
Interaction of Fock- and coherent-state wave packets with a two-level atom (2LA) in a 1D waveguide was studied within the Heisenberg picture in Refs.  \cite{dom,sto2013,sto2014,rou2016}.
 
A theoretical framework based on master equations was employed in Ref. \cite{bar} to describe the evolution of arbitrary quantum systems driven by $N$-photon wave packets. In Ref. \cite{tshi} a generalized master equation approach was developed for studying multiphoton transport in wQED systems with non-Markovian couplings.

Different quantum-field-theoretic approaches such as real-space Bethe ansatz \cite{shen2007,shen2007a,shen2015},
Lippmann-Schwinger formalism \cite{zheng}, Lehmann-Symanzik-Zimmermann reduction \cite{shi2009, shi2011, shi2013}, various diagrammatic techniques \cite{plet, schn, koc, see}, and Dyson series summation \cite{hurst} were adopted for the description of wQED systems as well.

An input-output formalism \cite{gard} was harnessed in Ref. \cite{sfan2010} to establish a relationship between the exact few-photon \textit{S} matrix of a 2LA and photonic input-output operators. This successful approach was employed in further studies \cite{koc2012, reph, xu2015, can, triv}.
 
It should be noted that the \textit{S} matrix provides information only about asymptotic (long-time) behavior of the system, while no insight into its transient dynamics is given. To describe the evolution of the system state one should rely on alternative techniques, e.g., wavefunction approach which was employed for studying the propagation of few-photon wave packets in wQED systems containing one or few 2LAs \cite{hof2003, ychen, nyst, konyk}, a single-photon pulse propagation in a chain of non-identical 2LAs \cite{liao2016}, interaction of one- \cite{wang} and two-photon pulses \cite{qhu} with a cavity-atom system, and few-photon scattering in systems with quantum delayed feedback \cite{fang2018, calajo}.
Detailed information about the system can be obtained if one knows its exact wavefunction at the arbitrary moment of time.
This can be achieved by solving a set of coupled equations of motion for probability amplitudes governing the quantum-state evolution of the system.
These first-order ordinary differential equations (ODEs) can be solved either numerically using one of the elaborate computational routines, or analytically, for some cases.

Many of the previous studies model a local emitter by a 2LA. It is the simplest saturable (i.e., nonlinear) quantum system since it can contain only one excitation per moment. This property was employed in Ref. \cite{rou} to derive an exact analytical solution for the problem of $N$-photon scattering on a 2LA.
Investigation of a multi-photon transport in wQED systems with more complex emitters constitutes a more intricate problem for analysis.
A system composed of a cavity coupled to an atom, which is referred to as the Jaynes-Cummings (JC) system, is one of the most important model systems in quantum optics. It demonstrates a number of interesting phenomena such as photon blockade \cite{birn, henn, hamsen}, photon phase switching \cite{tiecke}, generation of photon bundles \cite{sanch}, etc.

The paper concerns a few-photon transport in the wQED setup with the emitter represented by the JC system. 
We develop a wavefunction approach to provide a fully quantum-mechanical real-time analysis of scattering of two-photon Fock-state wave packets on the JC system coupled to a 1D unidirectional (chiral) waveguide.
The exact solutions of the equations of motion for the probability amplitudes determining a state of the system at the arbitrary moment of time are obtained both analytically and numerically.
In the long-time limit of the system evolution when all interaction processes are over, and the scattered photons propagate in the waveguide as free excitations, we ultimately arrive at the results equivalent to those derived earlier using other approaches.
We use Schmidt decomposition of the two-photon spectral distribution function (SDF) of the outgoing wave packet to demonstrate that the scattered photons are entangled.

The outline of the paper is as follows.
In Sec.~\ref{sec:model} a model system is described and its possible experimental incarnations are briefly discussed.
In Sec. \ref{sec:1phot} we study the case of a single-photon ingoing pulse. 
Sections \ref{sec:2phot}-\ref{sec:scatt} contain the results regarding the case of a two-photon Fock-state ingoing pulse.
In Sec. \ref{sec:2phot} the equations of motion governing the quantum-state evolution of the system are derived.
The dynamics of the JC system is studied in Section \ref{sec:jcdyn}.
In Sec. \ref{sec:scatt} we present the exact expression for the SDF of the wave packet scattered from the JC system.
The properties of the spectrum of the outgoing photons are analyzed in Sec. \ref{sec:spectr}.
In Sec. \ref{sec:entang} Schmidt decomposition is employed to investigate the frequency entanglement of the scattered photons.
We summarize in Sec.~\ref{sec:summ}. Details of derivation and solution of evolution equations are presented in Appendices.
\section{The Model} \label{sec:model}
\begin{figure}[t!] 
	\centering
	\includegraphics[width = 0.4\textwidth]{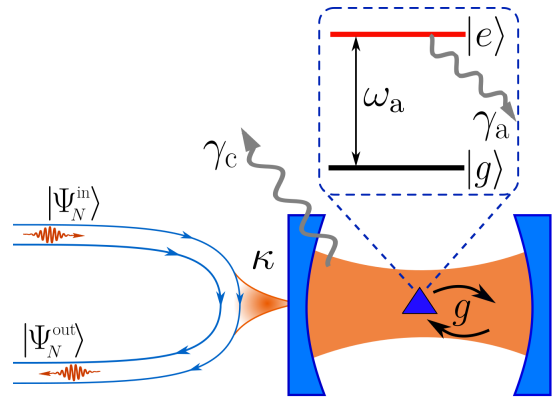}
	\caption{(Color online). A schematic illustration of the considered wQED setup. A chiral 1D waveguide is coupled to a JC system constituted by a cavity interacting with an atom with strength $\textstyle g$. Parameter $\kappa$ stands for a decay rate of the cavity mode caused by leakage into the waveguide.
	Both the cavity and the atom can have intrinsic dissipation rates $\textstyle \gamma_\mathrm{c}$ and $\textstyle \gamma_\mathrm{a}$, correspondingly. \label{fig:fig_1}} 
\end{figure}

We consider a model system consisting of a single-mode cavity coupled to an individual 2LA. A cavity mode overlaps with modes of a unidirectional 1D waveguide leading to that photons are able to leak in and out of the cavity. In order to populate the cavity and to excite the atom coupled to it, the former is driven by a few-photon wave packet propagating in a waveguide. The concept of the considered waveguide-cavity-atom system is illustrated in Fig. \hyperref[fig:fig_1]{\ref*{fig:fig_1}}.

In practice, the prototypical waveguide-cavity-atom system considered here can be implemented within a variety of physical architectures. Those include, but not limited to photonic-crystal structures forming waveguides and cavities for photons, which interact with natural atoms \cite{birn, tiecke}, individual semiconductor quantum dots (QDs) \cite{lohd2015}, or embedded color centers \cite{engl2010,radu2017}; tapered optical fibers coupled to whispering-gallery-mode resonators interacting with cold atoms trapped near their surface \cite{dayan}; superconducting circuit QED (cQED) setups, where nonlinear properties of the Josephson junctions (JJs) are utilized to embody atom-like systems \cite{clarke, schm, nori}.
Remarkably, but general properties of those diverse physical systems can be described by the same mathematical model discussed below.

\subsection*{The Hamiltonian}
The considered waveguide-cavity-atom system is described by the Hamiltonian \cite{shen, reph, oeh, chu2011}:
 \begin{equation} \label{eq:ham_1}
  \begin{split}
     \opH & = \overbrace{\omegac \, a^{\dag} a + \omegaa \, \sigma_+ \sigma_- + g \left(a^\dag \sigma_- + \sigma_+ a\right)}^{\opH_\mathrm{JC}} \\
     & \quad + \underbrace{\int^{\infty}_0 \dd \omega \, \omega \, b^{\dag}_{\omega} b_{\omega}}_{\opH_\mathrm{w}}
       + \underbrace{f \int^{\infty}_0 \dd \omega \left(b^{\dag}_\omega a + a^\dag b_\omega \right)}_{\opH_\mathrm{I}}.
  \end{split}
 \end{equation}
Henceforth, we set $\hbar \equiv 1$. Thus, throughout the paper, all energies are given in frequency units.

The first three terms on the right-hand side (rhs) of Eq. (\ref{eq:ham_1}) describe the interacting cavity-atom system and constitute the Hamiltonian of the JC model $\textstyle \opH_\mathrm{JC}$.
The first term in $\textstyle \opH_\mathrm{JC}$ describes a noninteracting single-mode cavity with a resonance frequency $\textstyle \omegac$, where $\textstyle a$ ($\textstyle a^{\dag}$) is a bosonic annihilation (creation) operator of a photon in the cavity mode. The second term in the Hamiltonian describes an individual 2LA with $\textstyle \omegaa$ being an energy gap between its ground $\textstyle |g\rangle$ and excited $|e\rangle$ states. We have introduced the atomic raising $\textstyle \sigma_+=|e\rangle\langle g|$ and lowering $\textstyle \sigma_- = |g\rangle\langle e|$ operators connected to the Pauli operators as $\textstyle \sigmax = \sigma_++\sigma_-$, $\textstyle \ii \sigmay = \sigma_+-\sigma_-$, and $\textstyle \sigmaz = 2\sigma_+\sigma_--1$. 
The Hamiltonian of the atom-cavity interaction, represented by the third term in the rhs of Eq. (1), is given in the rotating-wave approximation (RWA), which assumes the conditions
\begin{equation} \label{eq:rwa_crit}
 |\Deltaa| \ll \omegaa + \omegac, \quad g \ll \{\omegac, \omegaa\}
\end{equation}
are satisfied, where $\textstyle \Deltaa = \omegaa - \omegac$ denotes a detuning between frequencies of an atomic transition and a cavity resonance. Parameter $g$ stands for an atom-cavity coupling strength.
The criteria (\ref{eq:rwa_crit}) are satisfied for a broad spectrum of experimental wQED realizations ranging from microwave superconducting circuits to photonic-crystal structures operating at optical frequencies.
Nonetheless, one should note recent progress in the development of various system (see, e.g., reviews in Refs. \cite{pforndiaz, kockum} and references therein) exhibiting a light-matter coupling strong enough to break down the RWA applicability and switch the system into the ultrastrong coupling regime.

The term $\textstyle \opH_\mathrm{w}$ in Eq. (\ref{eq:ham_1}) describes a continuum of independent bosonic modes which models a waveguide.
Operator $\textstyle b_{\omega}$ ($\textstyle b^\dag_\omega$) annihilates (creates) a photon with energy $\textstyle \omega$ in the waveguide and obeys the equal-time commutation relation $\textstyle [b_{\omega}, b^{\dag}_{\omega'}] = \delta(\omega' - \omega)$.
The waveguide-cavity interaction is represented by the term $\textstyle \opH_\mathrm{I}$, where $\textstyle f \equiv \sqrt{\kappa/(2\pi)}$ is a waveguide-cavity coupling strength. Here we have neglected the actual frequency dependence of the cavity-waveguide coupling assuming that the frequencies of the cavity resonance and the incident photons are close to each other and much larger than the spectral bandwidth of the wave packet. This is expressed by the criteria
\begin{equation} \label{eq:narr_band}
 |\omega_0 - \omegac|\ll \omega_0 + \omegac, \quad \gamma_0 \ll \{\omega_0, \omegac\},
\end{equation}
where $\textstyle \omega_0$ is a central frequency of the incident pulse with a spectral bandwidth $\textstyle \gamma_0$.
Due to the second criterion in (\ref{eq:narr_band}) we extend boundaries of integration over the photon frequencies to $(-\infty,+\infty)$.
This approximation gives negligible deviation from the exact result due to the sharp localization of the wave packet spectrum in the vicinity of $\textstyle \omega_0$, which is expressed by the second criterion in (\ref{eq:narr_band}).

The model Hamiltonian contains no terms describing the interaction of the system with external reservoirs leading to dissipation, which implies that in the paper we are focused exclusively on a unitary evolution of the system.
This approximation is legitimate when the parameters of the system meet the constraint $\max\{\gamma_\mathrm{c},\gamma_\mathrm{a}\} \ll \min\{\gamma_0,\kappa\}$, which indicates that dissipation processes should be the slowest ones.
It is also assumed that the temperature $T_\mathrm{s}$ at which the system operates satisfies the criterion $\omegac/(k_\mathrm{B} T_\mathrm{s}) \gg 1 $, where $k_\mathrm{B}$ is the Boltzmann constant. This condition ensures that the average number of thermal excitations in the system is negligible $\textstyle n_\mathrm{th} = [\exp(-\frac{\omegac}{k_\mathrm{B} T_\mathrm{s}})-1]^{-1} \ll 1$.
The above conditions can be realized in the modern cQED systems (see Ref. \cite{schm} and references therein for the parameters of state-of-the-art superconducting cQED setups).

One can show that the operator of the total number of excitation in the system $\textstyle \hat{\mathcal{N}}_\mathrm{ex} = a^\dag a + \sigma_+\sigma_- + \int \dd \omega b^\dag_\omega b_\omega$ is an integral of motion since it commutes with the Hamiltonian: $\textstyle [\hat{\mathcal{N}}_\mathrm{ex}, \hat{\mathcal{H}}] = 0$, which indicates that the number of excitations in the system is conserved.

\section{Single-photon wave packet} \label{sec:1phot}
Let us first consider the case of the JC system driven by a single-photon wave packet.
As long as we set that initially the JC system contains no excitations (the atom resides in its ground state $|g\rangle$, and the cavity field is in the vacuum state), the evolution of the system occurs entirely within the one-excitation domain of the Hilbert space of the system states since the number of excitations in the system is conserved.
The time-dependent wavefunction of the system has the form
\begin{equation}  \label{eq:wavefunc1p}
   |\varPsi_1(t)\rangle = \left[A^g(t) a^\dag + A^e(t) \sigma_+ + \int \dd \omega B_{\omega}(t) b^\dag_\omega\right] \vacket,
\end{equation}
 where $\textstyle \vacket = \vacketw \vacketjc$ stands for a vacuum state of the entire system -- a state with no excitations in both the waveguide $\textstyle \vacketw$ and the JC system $\textstyle \vacketjc \equiv \vacketc|g\rangle$.
 In Eq. (\ref{eq:wavefunc1p}) $\textstyle B_{\omega}(t)$ is a single-photon SDF, $\textstyle A^g(t)$ stands for an amplitude of a state with the ground-state atom and the cavity field in the single-photon state, and $\textstyle A^e(t)$ is an amplitude of a state with the vacuum field in the cavity and the excited-state atom. Here and through the rest of the main part of the paper, we work within the Schr\"{o}dinger picture with time-independent operators.

 The initial (at $\textstyle t=0$) state of the system is given by
 \begin{equation}
  |\Psi^\mathrm{in}_1\rangle = |\psi^\mathrm{in}_1\rangle|\varnothing_\mathrm{JC}\rangle,
 \end{equation}
 where $\textstyle |\psi^\mathrm{in}_1\rangle$ is a state of the ingoing photon. It is given by
 \begin{equation}
   |\psi^\mathrm{in}_1\rangle = \int \dd \omega \, \xi_\omega \, b^\dag_\omega |\varnothing_\mathrm{w}\rangle,
  \end{equation}
 with $\textstyle \xi_\omega$ being an ingoing wave packet SDF satisfying a normalization condition $\textstyle \int \dd \omega \, |\xi_\omega|^2 = 1$.
 
 Evolution of the single-excitation amplitudes entering Eq. (\ref{eq:wavefunc1p}) is governed by a set of coupled equations as follows
 \begin{subequations} \label{eq:eqmot1}
 	\begin{gather}
 	 \left(\ii \partial_t - \tomegac\right) A^g(t) = f \Xi(t) + g A^e(t), \\
  	 \left(\ii\partial_t - \omegaa \right) A^e(t) = g A^g(t), \\
 	 \left(\ii \partial_t - \omega \right) B_\omega(t) = f A^g(t),  	 
 	\end{gather}
 \end{subequations}
 where $\textstyle \tomegac = \omegac - \ii\kappa/2$ and $\textstyle \Xi(t) = \int \dd \omega \, \ee^{-\ii \omega t} \, \xi_\omega$.
 The initial conditions are the following: $\textstyle B_\omega(0) = \xi_\omega$ and $\textstyle A^g(0)=A^e(0)=0$.
 The method of derivation of the above set of equations is presented in Appendix \ref{sec:der}.
 The exact analytical solutions of the system (\ref{eq:eqmot1}) can be obtained using the Laplace transform method (see the details in Appendix \ref{sec:sol}).
 
 Since the ingoing pulse has a finite duration $\tau_\mathrm{p} \sim \gamma_0^{-1}$ and the photons inside the cavity mode have limited lifetime $\sim \kappa^{-1}$, in the long-time limit
 \begin{equation} \label{eq:longtime}
  \textstyle t \gg \{\gamma_0^{-1}, \kappa^{-1}\} 
 \end{equation}
 the system reaches the steady state: the cavity field is in the vacuum state, the atom resides in its ground state, while the scattered photon propagates in the waveguide as a free excitation.
 Applying the condition (\ref{eq:longtime}) to solutions of Eq. (\ref{eq:eqmot1}) one obtains the final state of the system 
 $\textstyle |\Psi^\mathrm{out}_1\rangle = |\psi^\mathrm{out}_1\rangle \vacketjc$, 
 where
 \begin{equation} \label{eq:psi1_out}
  |\psi^\mathrm{out}_1\rangle = \int \dd \omega \, \ee^{-\ii \omega t} \, \ee^{\ii \varTheta_\omega} \xi_\omega \, b^\dag_\omega \vacketw,
 \end{equation}
 is the state of the scattered photon.
 Here $\textstyle \varTheta_\omega$ stands for a single-photon frequency-dependent phase shift induced by the JC system. 
 It is determined as \cite{shen, reph, oeh}:
 \begin{equation} \label{eq:theta}
  \varTheta_\omega = \mathrm{arg}\left[\frac{(\omega - \mathcal{E}^+_1)^*(\omega - \mathcal{E}^-_1)^*}{(\omega - \mathcal{E}^+_1)(\omega - \mathcal{E}^-_1)}\right].
 \end{equation}
 The complex single-photon resonances $\textstyle \mathcal{E}^\pm_1 = E^\pm_1 - \ii \varGamma^\pm_1/2$ of the \textit{open} (i.e. coupled to the waveguide) JC system are given by:
\begin{equation} \label{eq:epsilon1}
  \mathcal{E}^\pm_1 = \tomegac + \frac{\tDeltaa}{2} \pm \frac{\mathcal{R}_1}{2}, \quad \mathcal{R}_1 = \sqrt{4 g^2 + \tDeltaa^2},
\end{equation}
where $\textstyle \tDeltaa = \omegaa - \tomegac$. In the resonant case ($\textstyle \Deltaa=0$) one has
$\textstyle E^\pm_1 = \omegac \pm g \sqrt{1 - \kappa^2/(4g)^2}$ and $\textstyle \Gamma^\pm_1 = \kappa/2$.
For the strong-coupling regime of the JC system ($\textstyle g\gg\kappa$) one can employ an approximation $\textstyle E^\pm_1 \approx \omegac \pm g$.
\begin{figure*}[t!] 
	\centering
	\includegraphics[width = 0.98\textwidth]{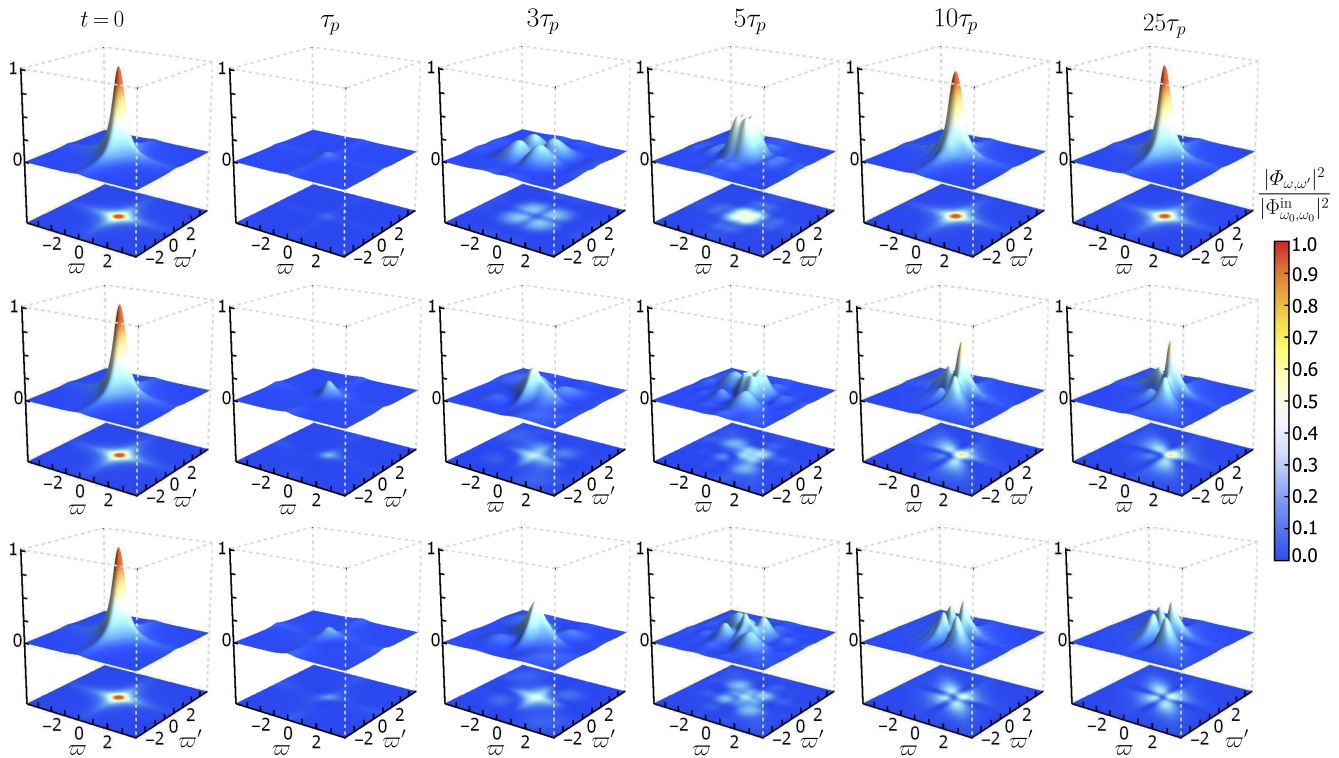}
	\caption{(Color online). Snapshots of evolution of the 2PJS in the process of scattering for specific moments of time and different parameters of the ingoing wave packet and the JC system. The JC system is considered in the resonant regime ($\textstyle \Deltaa = 0$) and is driven on one of the single-photon resonances $\textstyle \omega_0=E^+_1$.
	The rest of the parameters are as follows: $\textstyle \gamma_0/\kappa=0.2$, (upper row) $\textstyle g/\kappa=0$, $E^+_1=\omegac$;
	(middle row) $\textstyle g/\kappa=2$; (bottom row) $g/\kappa=10$.
	Here, for the sake of brevity, we use notations as follows: $\textstyle \varpi = (\omega-\omega_0)/\gamma_0$ and $\textstyle \varpi'=(\omega'-\omega_0)/\gamma_0$.
	All plots are normalized on the maximum of the ingoing wave packet 2PJS $\textstyle |\Phi^\mathrm{in}_{\omega_0,\omega_0}|^2 = 4/(\pi\gamma_0)^{2}$.
	See Ref. \cite{suppl} for the 2PJS evolution animation. \label{fig:fig_evol}} 
\end{figure*}
\section{Two-photon problem: Quantum-state evolution} \label{sec:2phot}
\subsection{The wavefunction}
Here we consider the evolution of the waveguide-JC system when the incident wave packet is in the two-photon Fock state.
Due to conservation of the excitation number the dynamics of the system is restricted exclusively by the two-excitation domain of the Hilbert space of the system states.
The wavefunction of the system at arbitrary moment $\textstyle |\varPsi_2(t)\rangle$ is represented by a superposition of two-excitation states:
 \begin{equation} \label{eq:wavefunc}
  \begin{split} 
    |\varPsi_2(t)\rangle & = \frac{1}{\sqrt{2}} \iint \dd \omega \, \dd \omega' \, \varPhi_{\omega,\omega'}(t) \, b^\dag_{\omega} b^\dag_{\omega'} \vacket \\
    & \quad + \int \dd \omega \left[X^{g}_{\omega}(t)  \, b^\dag_\omega a^\dag + X^{e}_{\omega}(t) \, b^\dag_\omega \sigma_+ \right]\vacket \\
    & \quad + \left[\frac{1}{\sqrt{2}} Z^g(t) \, \big(a^\dag\big)^2 + Z^e(t) \, \sigma_+ a^\dag \right] \vacket.
  \end{split}
 \end{equation}
The first term in Eq. (\ref{eq:wavefunc}) describes a state with both photons in the waveguide.
The time-dependent two-photon SDF $\textstyle \varPhi_{\omega,\omega'}(t)$ exhibits symmetry to permutation of photon frequencies $\textstyle \varPhi_{\omega,\omega'}(t) = \varPhi_{\omega',\omega}(t)$ because of the bosonic nature of photons.
A quantity $\textstyle |\varPhi_{\omega,\omega'}(t)|^2$ is referred to as a two-photon joint spectrum (2PJS) and determines a distribution of a joint probability of finding a pair of photons with frequencies $\textstyle \omega$ and $\textstyle \omega'$ in the waveguide.
The probability amplitudes $\textstyle X^{g,e}_\omega(t)$ correspond to the states of the system with the waveguide and the JC system both containing a single excitation. The amplitudes $\textstyle Z^{g,e}(t)$ correspond to the states with the waveguide void of photons and the JC system containing two excitations. Subscripts $\textstyle g$ and $\textstyle e$ specify the state of the atom (ground or excited). 

The initial state of the system in the case of the two-photon drive is $\textstyle |\Psi^\mathrm{in}_2\rangle = |\psi^\mathrm{in}_2\rangle\vacketjc$ with the wavefunction $\textstyle |\psi^\mathrm{in}_2\rangle$ of the ingoing photons given by
 \begin{equation} \label{eq:psi_0}
  |\psi^\mathrm{in}_2\rangle = \frac{1}{\sqrt{2}} \iint \dd \omega \, \dd \omega' \Phi^\mathrm{in}_{\omega,\omega'} \, b^\dag_{\omega} b^\dag_{\omega'} \vacketw.
 \end{equation}
In the above expression $\textstyle \Phi^\mathrm{in}_{\omega,\omega'}$ stands for a two-photon SDF of the incident wave packet obeying the condition $\textstyle \iint \dd \omega \dd \omega' \left|\Phi^\mathrm{in}_{\omega,\omega'}\right|^2 = 1$.

It is assumed that the incident two-photon wave packet is composed of a pair of indistinguishable photons with SDFs given by $\textstyle \xi_\omega$.
In this case, $\textstyle \Phi^\mathrm{in}_{\omega,\omega'}$ factorizes into the product of single-photon SDFs as
 \begin{equation} \label{eq:phi_in}
  \Phi^\mathrm{in}_{\omega,\omega'} = \xi_\omega \, \xi_{\omega'}.
 \end{equation}
Therefore, the initial photonic state $\textstyle |\psi^\mathrm{in}_2\rangle$ is separable and can be expressed in the form
 \begin{equation} \label{eq:init_state2}
  |\psi^\mathrm{in}_2\rangle = \frac{1}{\sqrt{2}} |\psi^\mathrm{in}_1\rangle |\psi^\mathrm{in}_1\rangle.
 \end{equation}
Thus, following the arguments by Rohde \textit{et. al} \cite{roh}, here we deal with an ingoing wave packet in the two-photon \textit{multimode Fock state}. In general, \textit{multimode Fock states} are a special case of a broader class of \textit{multimode multiphoton states}.
 
\subsection{Evolution equations}
Equation of motion governing the evolution of the two-photon SDF $\textstyle \varPhi_{\omega,\omega'}(t)$ reads as follows
 \begin{equation} \label{eq:eqmot_phi}
    \left[\ii\partial_t - (\omega + \omega')\right] \varPhi_{\omega,\omega'}(t) = \frac{f}{\sqrt{2}}\big[X^g_\omega(t) +  X^g_{\omega'}(t)\big],
 \end{equation}
 with the initial condition $\textstyle \varPhi_{\omega,\omega'}(0) = \Phi^\mathrm{in}_{\omega,\omega'}$.
 The rest of the probability amplitudes in Eq. (\ref{eq:wavefunc}) have zero initial conditions and obey the evolution equations as follows:
 \begin{subequations} \label{eq:eqmot2}
 \begin{equation}
  \begin{split}
   \left[\ii\partial_t - (\omega + \tomegac)\right] X^g_\omega(t) = & \, f \sqrt{2} \Xi(t) B_\omega(t) \\ & \, + g X^e_\omega(t) + f \sqrt{2} Z^g(t),
  \end{split}
  \end{equation}
  \begin{equation}
    \left[\ii\partial_t - (\omega + \omegaa)\right] X^e_{\omega}(t) = g X^g_\omega(t) + f Z^e(t),
  \end{equation}
  \begin{equation}
     \left[\ii\partial_t - 2\tomegac\right] Z^g(t) = 2 f \Xi(t) A^g(t) + g \sqrt{2} Z^e(t),
  \end{equation}
  \begin{equation}
   \left[\ii\partial_t - \left(\tomegac + \omegaa\right)\right] Z^e(t) = f \sqrt{2} \Xi(t) A^e(t) + g \sqrt{2} Z^g(t).
  \end{equation}  
 \end{subequations}
 The details of derivation of Eqs. (\ref{eq:eqmot_phi}) and (\ref{eq:eqmot2}) are given in Appendix \ref{sec:der}.
 We solve the above system of ODEs governing evolution of the amplitudes both numerically and analytically. For numerical solution we use \texttt{NDSolve} function of the \textit{Mathematica} system.
 The analytical solutions are obtained using the Laplace transform method (see derivations in Appendix \ref{sec:sol}).
 
 We proceed to investigation of evolution of the 2PJS of the wave packet in the course of its interaction with the JC system.
 For calculations we model the single-photon SDF of the ingoing wave packet $\textstyle \xi_\omega$ by the Lorentzian function given by
 \begin{equation} \label{eq:def_xi}
   \xi_\omega = \sqrt{\frac{\gamma_0}{2\pi}} \left[(\omega - \omega_0) + \ii \, \frac{\gamma_0}{2}\right]^{-1} \, \ee^{\ii (\omega - \omega_0) t_0}
 \end{equation}
 where parameter $\textstyle t_0$ denotes a moment of time when the incident wave packet arrives at the cavity site. Hereinafter we set $\textstyle t_0=0$.
 For $\textstyle \xi_\omega$ modeled by a Lorentzian shape determined by Eq. (\ref{eq:def_xi}) one obtains:
 \begin{equation} \label{eq:Xi_lor}
  \Xi(t) = - \sqrt{\frac{2\pi}{\tau_\mathrm{p}}} \, \exp\left[-\frac{t}{2\tau_\mathrm{p}}\right] \theta(t),
 \end{equation}
 where $\textstyle \theta(t)$ is the Heaviside function. 

 The animation of 2PJS evolution can be found in Supplemental Material \cite{suppl}, while the series of snapshots for the specific moments of time and different parameters of the system are demonstrated in Fig. \ref{fig:fig_evol}.
 These computations reveal that the 2PJS of the wave packet experiences a transformation in the course of its interaction with the JC system.
 Another important detail is that for $\textstyle g=0$, which corresponds to the setup with the cavity either empty or decoupled from the atom, the 2PJS ultimately regains its initial shape for times satisfying the condition (\ref{eq:longtime}).
 This result will be rigorously explained below in the paper.
 On the contrary, for the case of non-zero interaction ($\textstyle g>0$) between the cavity and the atom, the 2PJS of the scattered wave packets [for times satifying the criterion \eqref{eq:longtime}] are transformed compared to the initial 2PJS.
 As it will be demonstrated later in the paper (see Sec. \ref{sec:entang}), unlike the ingoing photons, which are not entangled [see Eq. (\ref{eq:psi_0})], for $\textstyle g>0$ the scattered photons are in the frequency-entangled state.
\section{JC system dynamics} \label{sec:jcdyn}
 For the analysis of the excitation dynamics of the JC system we use Eq. (\ref{eq:wavefunc}). The average photon number in the cavity is given by
 \begin{equation}
  \begin{split}
    N_\mathrm{c}(t) =  & \, \langle \varPsi_2(t)|a^\dag a|\varPsi_2(t)\rangle \\
    & \, = \int \dd \omega \, |X^g_\omega(t)|^2 + |Z^e(t)|^2 + 2 |Z^g(t)|^2.
  \end{split}
 \end{equation} 
 The atomic excited state population
 is determined as
 \begin{equation}
  \begin{split}
   P_\mathrm{a}(t) = & \, \langle \varPsi_2(t)|\sigma_+ \sigma_-|\varPsi_2(t)\rangle \\
   & \, = \int \dd \omega \, |X^e_\omega(t)|^2 + |Z^e(t)|^2.
  \end{split}
 \end{equation}

 \begin{figure}[t!] 
	\centering
	\includegraphics[width = 0.45\textwidth]{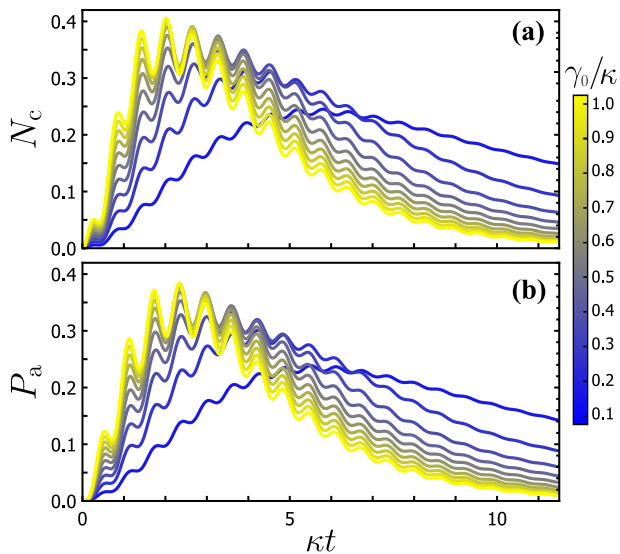}
	\caption{(Color online). Excitation dynamics of (a) the cavity and (b) the TLA for different values of $\textstyle \gamma_0/\kappa$, which are encoded by the color gradient. The parameters of the system are as follows: $\textstyle g/\kappa=5$, $\Deltaa=0$, and $\textstyle \omega_0=E^+_1$.
	\label{fig:fig_dynamics}}
 \end{figure} 
 
 Moreover, using the exact wavefunction (\ref{eq:wavefunc}) one can calculate the probability to find the JC system in the state with a given number of excitations $\textstyle p^{(j)}_\mathrm{JC}(t)$ ($\textstyle j\leq 2$). The probabilities of finding one and two excitations in the JC system at an instant $\textstyle t$ are determined as 
 \begin{equation}
  p^{(1)}_\mathrm{JC}(t) = \int \dd\omega |X^g_\omega(t)|^2 + \int \dd \omega |X^e_\omega(t)|^2
 \end{equation}
 and
 \begin{equation}
  p^{(2)}_\mathrm{JC}(t) = |Z^g(t)|^2 + |Z^e(t)|^2.
 \end{equation} 
 \begin{figure}[t!] 
	\centering
	\includegraphics[width = 0.45\textwidth]{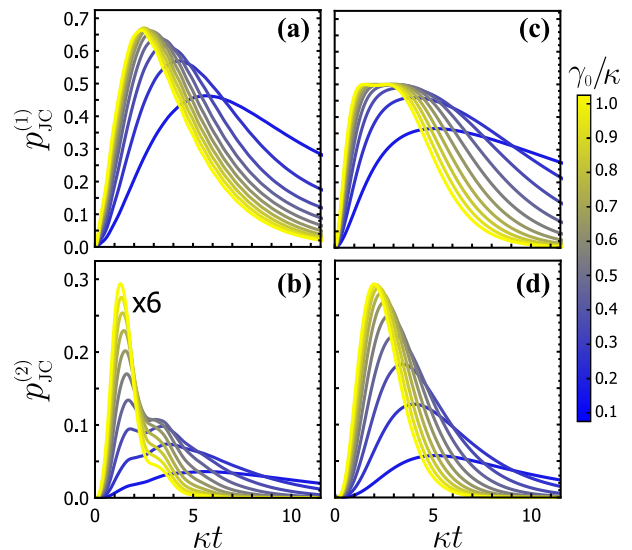}
	\caption{(Color online). Figures (a) and (b) show the dependences of the probabilities of finding one and two excitations in the JC system on $\textstyle \gamma_0/\kappa$ for $\textstyle g/\kappa=5$. The two-excitation probabilities on figure (b) are plotted after being multiplied by $\textstyle 6$. Figures (c) and (d) show the same dependences for the case of the empty cavity ($\textstyle g=0$).
	The other parameters and notations are as in Fig. \ref{fig:fig_dynamics}. \label{fig:fig_prob}}
\end{figure}

 Calculations of the cavity and atomic excitation dynamics presented in Fig. \ref{fig:fig_dynamics} show that shorter ingoing pulses (larger ratio $\textstyle \gamma_0/\kappa$) results in more efficient excitation of both the cavity and the 2LA.
 The cavity and atomic populations exhibit the oscillations arising due to exchange of excitations between the 2LA and the cavity.
 For times exceeding the duration of the ingoing pulse $\textstyle t > \tau_\mathrm{p}$ the cavity and atomic populations decay exponentially as $\propto \ee^{-\kappa t}$.
 
 Computations of $\textstyle p^{(1)}_\mathrm{JC}$ and $\textstyle p^{(2)}_\mathrm{JC}$ shown in Fig. \ref{fig:fig_prob} reveal that the probability of finding the JC system in the two-excitation state is considerably reduced compared to the empty cavity ($\textstyle g=0$) case (see Figs. \ref{fig:fig_prob}c and \ref{fig:fig_prob}d). This is explained as follows. Coupling the cavity mode to the 2LA, which is a saturable (i.e. nonlinear) system due to its ability to emit or absorb only one photon at a moment, introduces nonlinearity into the composite JC system. This nonlinearity gives rise to the photon blockade effect \cite{birn, henn} when the presence of one excitation in the system prevents the appearance of the second one.
 
 Since the probability of the JC system to reside in the two-excitation state is inhibited due to the atom-induced photon blockade, the frequency of oscillations of the cavity and 2LA populations is approximately the vacuum Rabi frequency of the JC system $\textstyle \sqrt{g^2 + (\Deltaa/2)^2}$.
 A slight mismatch between these frequencies is due to the contribution of two-excitation processes and cavity-waveguide coupling.
\section{Scattered two-photon state} \label{sec:scatt}
 Besides the transient quantum-state dynamics of the system, the asymptotic (long-time) solutions are of considerable interest as well.
 For times satisfying the condition (\ref{eq:longtime}), one arrives at the result that only the term which corresponds to the free evolution of the scattered two-photon field in the waveguide survives in the wavefunction of the system expressed by Eq. (\ref{eq:wavefunc}).  
 The remaining terms vanish indicating that the JC system is completely depopulated and rests in its ground state $\textstyle \vacketjc \equiv \vacketc |g\rangle$. The final state of the system is then given by $\textstyle |\Psi^\mathrm{out}_{2}\rangle = |\psi^\mathrm{out}_2\rangle\vacketjc$, where $\textstyle |\psi^\mathrm{out}_2\rangle$ stands for the scattered photons wavefunction. It is expressed as
 \begin{equation} \label{eq:wavefunc_out}
  |\psi^\mathrm{out}_{2}\rangle = \frac{1}{\sqrt{2}} \iint \dd \omega \, \dd \omega' \, \ee^{-\ii(\omega+\omega') t} \Phi^\mathrm{out}_{\omega,\omega'} \, b^\dag_{\omega} b^\dag_{\omega'} \vacketw,
 \end{equation} 
 where $\textstyle \Phi^\mathrm{out}_{\omega,\omega'}$ is the two-photon SDF of the scattered wave packet given by (see derivation in Appendix \ref{sec:sol}):
 \begin{equation} \label{eq:phiout}
  \begin{split}
     \Phi^\mathrm{out}_{\omega,\omega'} = \overbrace{\ee^{\ii (\varTheta_\omega + \varTheta_{\omega'})} \xi_\omega \xi_{\omega'}}^{\phi^\mathrm{el}_{\omega,\omega'}} + \overbrace{F_{\omega,\omega'} \prod_{\mu=\pm}\xi_{\omega+\omega'-\mathcal{E}^\mu_1}}^{\phi^\mathrm{inel}_{\omega,\omega'}},
  \end{split}
 \end{equation}
 where $\textstyle F_{\omega,\omega'}$ has the form
 \begin{equation}
 \begin{split}
   F_{\omega,\omega'} = & \, 16 \pi^2 f^4 g^4 \left[\frac{1}{2}+\frac{\omega+\omega'-\mathcal{E}^+_1-\mathcal{E}^-_1}{\omega+\omega'+\ii\gamma}\right] \\
   & \, \times \prod_{\mu=\pm}\frac{1}{(\omega - \mathcal{E}^\mu_1)(\omega' - \mathcal{E}^\mu_1)(\omega + \omega' - \mathcal{E}^\mu_2)}.
 \end{split}
 \end{equation}
 The result expressed by Eq. (\ref{eq:phiout}) can be equivalently derived using the two-photon \textit{S} matrix of the JC system obtained, e.g., in Refs. \cite{shi2011, reph, oeh}. 
 
 The term $\textstyle \phi^\mathrm{el}_{\omega,\omega'}$ in Eq. (\ref{eq:phiout}) corresponds to the uncorrelated elastic scattering of two photons when the energy of each one is conserved.
 The term $\textstyle \phi^\mathrm{inel}_{\omega,\omega'}$ describes inelastic scattering when the energies of individual photons are not conserved.
 The complex two-photon resonances of the open JC system $\textstyle \mathcal{E}^\pm_2 = E^\pm_2 - \ii \Gamma^\pm_2/2$ are given by
 \begin{equation} \label{eq:lambda2}
  \mathcal{E}^\pm_2 = 2\tomegac + \frac{\tDeltaa}{2} \pm \frac{\tR_2}{2}, \quad \tR_2 = \sqrt{8 g^2+\tDeltaa^2}.
 \end{equation} 
\subsection*{Cavity decoupled from atom}
 Let us consider a limiting case of the JC system when the cavity is empty or decoupled from the atom ($\textstyle g=0$).
 In this case, the inelastic term $\textstyle \phi^\mathrm{inel}_{\omega,\omega'}$ vanishes in Eq. (\ref{eq:phiout}), and the two-photon SDF of the scattered wave packet reduces to
\begin{equation} \label{eq:phioutg0}
 \Phi^\mathrm{out}_{\omega,\omega'} \overset{g=0}{\rightarrow} \mathring{\Phi}^\mathrm{out}_{\omega,\omega'} = \mathring{\xi}_\omega \, \mathring{\xi}_{\omega'},
\end{equation}
where $\textstyle \mathring{\xi}_\omega = \exp(\ii \mathring{\varTheta}_\omega) \, \xi_\omega$ is the SDF of the scattered photon with the single-photon phase shift given by
$\textstyle \mathring{\varTheta}_\omega=\mathrm{arg}\left[(\omega-\tilde{\omega}^*_\mathrm{c})/(\omega-\tilde{\omega}_\mathrm{c})\right]$.
Thereby, the photons acquire only phase shifts in the course of scattering from the cavity which does not contain an atom.
Using Eq. (\ref{eq:wavefunc_out}) one has
\begin{equation} \label{eq:psioutg0}
 |\mathring{\psi}^\mathrm{out}_2\rangle = |\mathring{\psi}^\mathrm{out}_1\rangle |\mathring{\psi}^\mathrm{out}_1\rangle,
\end{equation}
where $\textstyle |\mathring{\psi}^\mathrm{out}_1\rangle = \int \dd \omega \, \mathring{\xi}_\omega b^\dag_\omega \vacketw$.
The above expression indicates that for $\textstyle g=0$ the state of the scattered photons is separable.

Equation (\ref{eq:phioutg0}) implies that the 2PJS of the outgoing wave packet is not transformed compared to that of the ingoing one:
\begin{equation} \label{eq:2pjs0}
 |\mathring{\Phi}^\mathrm{out}_{\omega, \omega'}|^2 = |\Phi^\mathrm{in}_{\omega,\omega'}|^2,
\end{equation}
which supports the results presented in Fig. \ref{fig:fig_evol} regarding the empty cavity case.
\section{Spectrum of scattered photons} \label{sec:spectr}
\begin{figure} [t!]
	\centering
	\includegraphics[width = 0.45\textwidth]{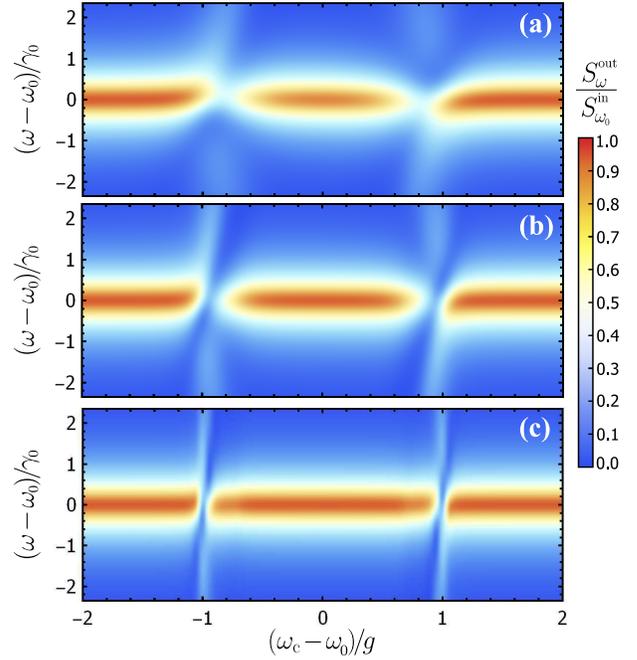}
	\caption{(Color online). Dependence of the spectrum of the scattered wave packet on $\textstyle \omega_0$ for the JC system in the resonant regime ($\textstyle \Deltaa=0$, $\textstyle E^\pm_1 = \omegac \pm g \sqrt{1 - \kappa^2/(4g)^2}$), 
		$\textstyle \gamma_0/\kappa=0.2$, and (a) $\textstyle g/\kappa=1$, (b) $\textstyle g/\kappa=2$, (c) $\textstyle g/\kappa=10$. \label{fig:fig_spectra}}
\end{figure}
 Spectrum of the scattered photons is determined as $\textstyle S^\mathrm{out}_\omega = \langle \Psi^\mathrm{out}_2|b^\dagger_\omega b_\omega|\Psi^\mathrm{out}_2\rangle$ \cite{sto2013,sto2014}, and characterizes the density of photons with frequency $\textstyle \omega$ in the scattered wave packet. The average number of photons in the frequency band $\textstyle \left[\Omega, \Omega+\delta\right]$ is given by $\textstyle \int^{\Omega+\delta}_{\Omega} \dd \omega S^\mathrm{out}_\omega$. Integration of $\textstyle S^\mathrm{out}_\omega$ over the entire frequency range gives the average scattered photon number.
 Since the number of excitation in the system is conserved, one has $\textstyle \int \dd \omega \, S^\mathrm{out}_\omega = \int \dd \omega \, S^\mathrm{in}_\omega = 2$, where $\textstyle S^\mathrm{in}_\omega = \langle \Psi^\mathrm{in}_2|b^\dag_\omega b_\omega|\Psi^\mathrm{in}_2\rangle = 2 |\xi_\omega|^2$ is the spectrum of the ingoing wave packet.
 Using Eq. (\ref{eq:wavefunc_out}) one arrives at the expression for $\textstyle S^\mathrm{out}_\omega$ as follows:
\begin{equation} \label{eq:spectr}
 S^\mathrm{out}_\omega = 2 \int \dd \omega' \left|\Phi^\mathrm{out}_{\omega, \omega'}\right|^2.
\end{equation}
It follows from Eqs. (\ref{eq:spectr}) and (\ref{eq:2pjs0}) that for the setup with the cavity decoupled from the atom one has $\textstyle S^\mathrm{out}_\omega = S^\mathrm{in}_\omega$ implying that for $\textstyle g=0$ the spectrum of the scattered photons is identical to that of the ingoing ones.

\begin{figure} [t!]
	\centering
	\includegraphics[width = 0.45\textwidth]{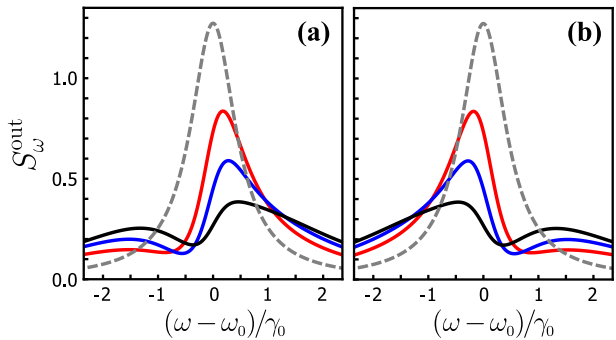}
	\caption{(Color online). Spectra of the scattered wave packets for (a) $\textstyle \omega_0 = E^+_1$ and (b) $\textstyle \omega_0 = E^-_1$, and different atom-cavity coupling strengths: (red line) $\textstyle g/\kappa=1$, (blue line) $\textstyle g/\kappa=2$, and (black line) $\textstyle g/\kappa=10$. The rest of the parameters are the following: $\textstyle \gamma_0/\kappa=0.2$ and $\textstyle \Deltaa=0$. Dashed lines demonstrate the spectrum of the ingoing wave packet $\textstyle S^\mathrm{in}_\omega = 2 |\xi_\omega|^2$. \label{fig:fig_spectra2}}
\end{figure}

Figure \ref{fig:fig_spectra} demonstrates the dependence of the scattered photons spectrum on the central frequency of the ingoing wave packet $\textstyle \omega_0$. Calculations reveal that the most pronounced modification of the scattered wave packet spectrum occurs when $\textstyle \omega_0$ lies in the vicinity of the JC single-photon resonances $\textstyle \omega_0\approx E^\pm_1$. The effect of atom-cavity coupling strength $\textstyle g$ on the scattered spectrum for $\textstyle \omega_0=E^\pm_1$ is shown in Fig. \ref{fig:fig_spectra2}. One can see that increased atom-cavity coupling leads to stronger modification of the scattered photons spectrum.

\begin{figure} [t!]
	\centering
	\includegraphics[width = 0.45\textwidth]{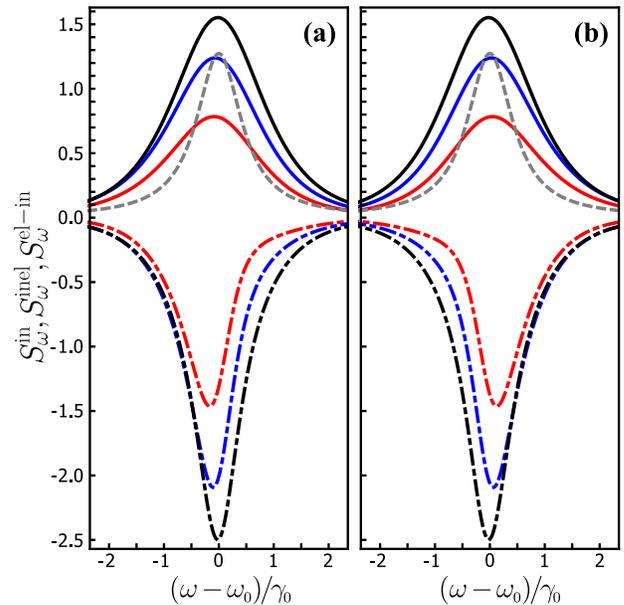}
	\caption{(Color online). Plots of different contributions to the outgoing photons spectrum: $\textstyle S^\mathrm{inel}_\omega$ (solid lines),   $\textstyle $ (dash-dotted lines) and $\textstyle S^\mathrm{in}_\omega$ (gray dashed line) for (a) $\textstyle \omega_0 = E^+_1$ and (b) $\textstyle \omega_0 = E^-_1$.
		Other notations are as in Fig. \ref{fig:fig_spectra2}.
		\label{fig:fig_spectra3}}
\end{figure}

If one substitutes Eq. (\ref{eq:phiout}) into Eq. (\ref{eq:spectr}) the scattered photons spectrum $\textstyle S^\mathrm{out}_\omega$ can be represented as
\begin{equation}
 S^\mathrm{out}_\omega = S^\mathrm{in}_\omega + S^\mathrm{inel}_\omega + S^\mathrm{el-in}_\omega.
\end{equation}
In the above expression $\textstyle S^\mathrm{inel}_\omega$ stands for the contribution to the scattered photons spectrum arising purely from the inelastic scattering determined by
\begin{equation}
 S^\mathrm{inel}_\omega = 2 \int \dd \omega' |\phi^\mathrm{inel}_{\omega,\omega'}|^2.
\end{equation}
The term
\begin{equation}
 S^\mathrm{el-in}_\omega = 4 \int \dd \omega' \, \mathrm{Re}\left\{(\phi^\mathrm{el}_{\omega,\omega'})^* \phi^\mathrm{inel}_{\omega,\omega'}\right\}
\end{equation}
arises due to interference of components of the outgoing wave packet produced by elastic and inelastic scattering from the JC system.
Contributions of all these terms to the scattered photons spectrum for the different parameters of the system are shown in Fig. \ref{fig:fig_spectra3}. Calculations reveal that $\textstyle S^\mathrm{el-in}_\omega$ takes negative values producing the ``dip" in the shape of the scattered photons spectrum seen in Fig. \ref{fig:fig_spectra2}. 
\section{Frequency entanglement} \label{sec:entang}
As it was rigorously shown in the Sec. \ref{sec:scatt} when the two-photon wave packet interacts with the cavity decoupled from the atom the scattered photons acquire only phase shifts while their SDF remains factorable. The state of the scattered photons is separable as that of the ingoing ones.
The situation changes starkly when one switches on the coupling between the cavity and the atom.
The JC nonlinearity gives rise to the effective photon-photon interaction which results in strongly-correlated scattering (photon blockade) discussed in Sec. \ref{sec:jcdyn}. According to Ref. \cite{xu2013}, the photon-photon interaction leads to frequency entanglement.
In our setup the latter arises due to the presence of the ``inelastic" term $\textstyle \phi^\mathrm{inel}_{\omega,\omega'}$ in the SDF of the scattered wave packet.

The frequency entanglement of the scattered photons can be quantified using the Schmidt decomposition of their SDF \cite{fed}:
\begin{equation} \label{eq:decomp}
\Phi^\mathrm{out}_{\omega, \omega'} = \sum_{j\geq 1} \sqrt{\lambda_j} \, \varphi_{j, \omega} \widetilde{\varphi}_{j, \omega'},
\end{equation}
which represents the SDF of the outgoing two-photon wave packet as a weighted sum of products of single-photon SDFs.
The decomposition coefficients satisfy $\textstyle \sum_j \lambda_j=1$.
The Schmidt-mode single-photon SDFs $\textstyle \varphi_{j, \omega}$ and $\textstyle \widetilde{\varphi}_{j,\omega}$ form a complete set of orthonormal functions: $\textstyle \int \dd \omega \, \varphi^*_{i,\omega} \widetilde{\varphi}_{j,\omega} = \delta_{i,j}$ and $\textstyle \sum_{i} \varphi^*_{i, \omega'} \widetilde{\varphi}_{i,\omega} = \delta(\omega'-\omega)$. The Schmidt coefficients and SDFs are determined via solution of the eigenvalue problem \cite{par,wal}:
\begin{equation} \label{eq:eigprobl}
\begin{split}
 \int \dd\nu \, K_{\nu,\omega} \varphi_{j,\nu} = \lambda_j \varphi_{j,\omega}, \\
 \int \dd\nu \, \widetilde{K}_{\nu,\omega} \widetilde{\varphi}_{j,\nu} = \lambda_j \widetilde{\varphi}_{j,\omega}, \\ 
\end{split}
\end{equation}
where the integral kernels $\textstyle K_{\nu,\omega}$ and $\textstyle \widetilde{K}_{\nu,\omega}$ are given by
\begin{equation}
 \begin{split}
   K_{\nu,\omega} = \int \dd\nu' \left(\Phi^{\mathrm{out}}_{\nu,\nu'}\right)^* \Phi^\mathrm{out}_{\omega,\nu'}, \\
  \widetilde{K}_{\nu,\omega} = \int \dd\nu' \left(\Phi^\mathrm{out}_{\nu',\nu}\right)^* \Phi^\mathrm{out}_{\nu',\omega}.
 \end{split}
\end{equation}
Since the SDF obeys the permutation symmetry $\textstyle \Phi^\mathrm{out}_{\nu,\nu'} \equiv \Phi^\mathrm{out}_{\nu',\nu}$, this leads to $\textstyle K_{\nu,\omega} = \widetilde{K}_{\nu,\omega}$ which, in turn, gives $\textstyle \varphi_{j,\omega} = \widetilde{\varphi}_{j,\omega}$.
We tackle the eigenvalue problem (\ref{eq:eigprobl}) numerically. For this purpose we discretize $\textstyle K_{\nu,\omega}$ into a uniform $\textstyle 100 \times 100$ grid on a square domain of frequencies ranging $\textstyle \pm 25\gamma_0$ around $\textstyle \omega_0$. 

\begin{figure}[t!]
	\centering
	\includegraphics[width = 0.45\textwidth]{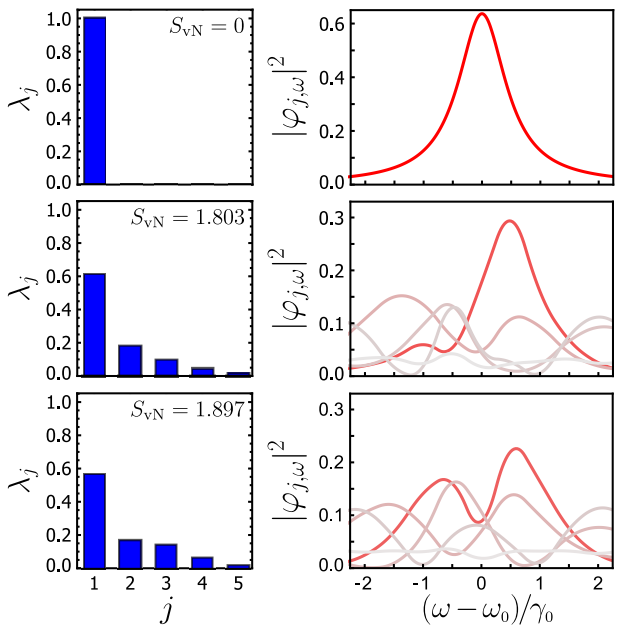}
	\caption{(Color online). First five Schmidt coefficients and Schmidt-mode joint spectra $\textstyle |\varphi_{j,\omega}|^2$ (saturation of lines encodes the value of the corresponding Schmidt coefficient) for the empty cavity $\textstyle g/\kappa=0$ (upper row), $\textstyle g/\kappa=2$ (middle row), and $\textstyle g/\kappa=10$ (bottom row). The rest of the parameters are identical to those used in Fig. \ref{fig:fig_evol}. \label{fig:fig_ent}}
\end{figure}

Using the decomposition (\ref{eq:decomp}) along with Eq. (\ref{eq:wavefunc_out}) one represents the outgoing photons state $\textstyle |\psi^\mathrm{out}_2\rangle$ as a superposition of two-photon separable states:
\begin{equation}
 |\psi^\mathrm{out}_2\rangle = \frac{1}{\sqrt{2}}\sum_{j\geq 1}\sqrt{\lambda_j}|\varphi_j\rangle |\varphi_j\rangle,
\end{equation}
where $\textstyle |\varphi_j\rangle = \int \dd \omega \, \varphi_{j,\omega} a^\dag_\omega \vacketw$.

If Schmidt coefficients are all zero except for one, this indicates that the two-photon SDF is factorable and the photons are not entangled.
Otherwise, we deal with the frequency-entangled photon pair.
As a measure of entanglement, we also employ the von Neumann (entanglement) entropy $\textstyle S_\mathrm{vN}$.
For entangled states $\textstyle S_\mathrm{vN} > 0$, while zero entanglement entropy $\textstyle S_\mathrm{vN} = 0$ indicates that the quantum state is separable.
The von Neumann entropy is expressed in terms of Schmidt coefficients as \cite{benn}
\begin{equation} \label{eq:s_vn}
 S_\mathrm{vN} = - \sum_{j \geq 1} \lambda_j \log_2 \lambda_j.
\end{equation}

It follows from Eq. (\ref{eq:phioutg0}) that for the setup with the empty cavity ($\textstyle g=0$) the SDF of the outgoing photons is factorable and there is only one nonzero Schmidt coefficient $\textstyle \lambda_1 = 1$ indicating that there is no frequency entanglement.
On the contrary, for the cavity interacting with the atom ($\textstyle g>0$) the outgoing photons exhibit frequency entanglement since $\textstyle \lambda_{j\geq 2} > 0$ and $\textstyle S_\mathrm{vN} > 0$ as presented in Fig. \ref{fig:fig_ent}.
\section{Summary} \label{sec:summ} 
 In this section, we summarize our results.
 We studied the scattering of two-photon Fock-state wave packets on the JC system employing a wavefunction approach.
 Using the interrelation between the Heisenberg and Schr\"{o}dinger representations (see Appendix \ref{sec:der}) we derived sets of coupled equations of motions governing the evolution of the probability amplitudes which determine the state of the system at the arbitrary moment of time.
 Using the exact solutions of these equations of motion we tracked the evolution of a 2PJS of a wave packet in the process of its interaction with the JC system. The excitation dynamics of the latter was studies as well. We determined the probabilities to find the JC system containing one and two excitations at the arbitrary instant. We found that the probability of finding two excitations in the JC system is inhibited compared to the case of the empty cavity which indicates the photon blockade induced by the 2LA nonlinearity.
 
 In the long-time limit, when the system reaches its steady state, we derived the exact expressions for the SDF of the scattered two-photon wave packet. This result agrees with that obtained using the exact \textit{S} matrix of the JC system derived in the literature.
 We also analyzed the scattered photons spectrum. Our calculations reveal that it differs significantly from the incident photons spectrum when the central frequency of the ingoing two-photon wave packet is close to one of the JC single-photon resonances.
  
 We employed Schmidt decomposition and the von Neumann entropy as a measure of entanglement of the outgoing photons.
 We found that the photons scattered from the JC constitute a frequency-entangled photon pair or a biphoton.
 The waveguide-JC system can be exploited as a tunable deterministic source of itinerant frequency-entangled biphotons.
 This source can be built on the state-of-the-art superconducting cQED architecture.

 In the paper, we focus on the case of the ingoing photons without spectral entanglement, thereby, investigation of the local quantum system response to the frequency-entangled few-photon states is of interest.
 The approach employed in the paper can be used for studying more complex quantum systems, for example, multilevel emitters, coupled-resonator arrays, optomechanical systems. Moreover, one can go beyond the RWA employed in the paper for the description of atom-cavity coupling and consider multi-photon scattering on a quantum system operating in the regime of ultrastrong light-matter interaction.
 Consideration of those systems and regimes constitute possible directions for the follow-up studies.
\begin{acknowledgments}
 The author thanks O. O. Chumak for reading of the manuscript and useful comments, and A. Sokolov and S. Lukyanets for fruitful discussions.
\end{acknowledgments}
\appendix
\section{Derivation of evolution equations for the probability amplitudes} \label{sec:der}
 Here we present derivations of the equations of motion which govern the evolution of the single- and two-excitation probability amplitudes.
 For that purpose, in this Appendix we work within the Heisenberg picture and start with the derivation of the evolution equations for the operators introduced earlier in the paper. The model Hamiltonian given by Eq. (\ref{eq:ham_1}) generates the equation of motion for the waveguide variable $\textstyle b_\omega$ as follows
 \begin{equation} \label{eq:eqmot_a}
 \ii \partial_t \, b_{\omega} = \left[b_{\omega}, \opH\right] = \omega b_{\omega} + f a,
 \end{equation}
 which can be formally integrated as
 \begin{equation} \label{eq:form_sln}
  b_{\omega}(t) = \tilde{b}_{\omega}(t) - \ii f \int^t_{0} \dd \tau \, \ee^{-\ii \, \omega \, (t - \tau)} \, a(\tau).
 \end{equation}
 where $\tilde{b}_{\omega}(t) = b_\omega(0) \ee^{-\ii \omega t}$ stands for the annihilation operator of a photon propagating in the waveguide as a free excitation.
 
 Equation of motion for the cavity photon annihilation operator reads as
 \begin{equation} \label{eq:eqmot_c1}
  \left(\ii \partial_t - \omegac\right) c =  g \sigma_{-} + f \int \dd \omega \, b_\omega.
 \end{equation}
 Eliminating the reservoir variables by substituting Eq. (\ref{eq:form_sln}) into Eq. (\ref{eq:eqmot_c1}) and integrating over $\textstyle \omega$ one arrives at the equation of the form
 \begin{equation} \label{eq:eqmot_c}
  \left(\ii \partial_t -\tomegac\right) a = g \sigma_- + f \opB.
 \end{equation}
 where we have introduced an operator $\textstyle \opB(t) = \int \dd \omega \, \tilde{b}_\omega(t)$.
 Using Eq. (\ref{eq:form_sln}) one can show that the commutation relations hold
 \begin{equation} \label{eq:comm}
  \left[\opB, b_\omega\right] = \left[\opB, a\right] = \left[\opB, \sigma_-\right] = 0.
 \end{equation}
 These commutators are routinely utilized below for derivation of equations of motion for the probability amplitudes.
 
 The Heisenberg equation for the atomic lowering operator $\sigma_-$ takes the form 
 \begin{equation} \label{eq:eqmot_qub}
  \left(\ii \partial_t - \omegaa\right) \sigma_- = - g \sigmaz a.
 \end{equation}
 Equations of motion for the creation operators are derived by the Hermitian conjugation of the corresponding equations for the annihilation operators.

 Using Eq. (\ref{eq:wavefunc1p}) one can formally express the single-excitation probability amplitudes in the form as follows:
 \begin{equation}
  \begin{split}
   A^g(t) & = \vacbra a(0)|\varPsi_1(t)\rangle, \\
   A^e(t) & = \vacbra \sigma_-(0)|\varPsi_1(t)\rangle, \\
   B_\omega(t) & = \vacbra b_\omega(0)|\varPsi_1(t)\rangle.
  \end{split}
 \end{equation}
 Employing the property of the vacuum state $\textstyle \ee^{-\ii \opH t}\vacket = \vacket$ along with the relations $\textstyle |\varPsi_1(t)\rangle = \ee^{-\ii \opH t}|\Psi^\mathrm{in}_{1}\rangle$ and $\textstyle \opO(t) = \ee^{\ii \opH t} \opO(0) \ee^{-\ii \opH t}$, where $\textstyle \opO(t)$ denotes some quantum-mechanical operator, one can transfer the time dependence from the wavefunction onto the operators which gives $\textstyle \vacbra \opO(0)|\varPsi_1(t)\rangle = \vacbra \opO(t)|\Psi^\mathrm{in}_1\rangle$.
 By doing so, one obtains the following representation for the probability amplitudes:
 \begin{equation}
  \begin{split}
    A^g(t) & = \vacbra a(t)|\Psi^\mathrm{in}_1\rangle, \\
    A^e(t) & = \vacbra \sigma_-(t)|\Psi^\mathrm{in}_1\rangle, \\
    B_\omega(t) & =  \vacbra b_\omega(t)|\Psi^\mathrm{in}_1\rangle.
  \end{split}
 \end{equation}
 Using this representation along with the Heisenberg equations for the corresponding operators and taking into account that $\textstyle \langle\varnothing|\sigmaz = - \langle\varnothing|$ and $\textstyle \opB(t)|\Psi^\mathrm{in}_1\rangle = \Xi(t)\vacket$ one immediately arrives at the set of equations of motion (\ref{eq:eqmot1}).
 
 Following the same receipt, we represent the two-excitation amplitudes entering the wavefunction (\ref{eq:wavefunc}) as
 \begin{equation}
  \begin{split}
   \varPhi_{\omega,\omega'}(t) & = \frac{1}{\sqrt{2}}\vacbra b_\omega(t) b_{\omega'}(t)|\Psi^\mathrm{in}_2\rangle, \\
   X^g_\omega(t) & = \vacbra b_\omega(t)a(t)|\Psi^\mathrm{in}_2\rangle, \\ 
   X^e_\omega(t) & = \vacbra b_\omega(t)\sigma_-(t)|\Psi^\mathrm{in}_2\rangle, \\ 
   Z^g(t) & = \frac{1}{\sqrt{2}}\vacbra a^2(t)|\Psi^\mathrm{in}_2\rangle, \\ 
   Z^e(t) & = \vacbra \sigma_-(t) a(t)|\Psi^\mathrm{in}_2\rangle.
  \end{split}
 \end{equation}
 Employing the Heisenberg equations (\ref{eq:eqmot_a}), (\ref{eq:eqmot_c}) and (\ref{eq:eqmot_qub}), the commutators (\ref{eq:comm}) along with the properties $\textstyle \vacbra \sigma_+ = 0$ and $\textstyle \opB|\Psi^\mathrm{in}_2\rangle = \sqrt{2} \Xi(t) |\Psi^\mathrm{in}_1\rangle$ one obtains equations of motion for the two-excitation amplitudes (\ref{eq:eqmot2}).
 
 This scheme can be directly applied to the $\textstyle N$-excitation problem.
 When the JC system is driven by the $\textstyle N$-photon Fock-state wave packet, one arrives at the set of coupled equations of motions, where $N$-excitation amplitudes depend on $\textstyle (N-1)$-excitation amplitudes and so on down to $\textstyle 1$-excitation amplitudes governed by Eq. (\ref{eq:eqmot1}). This hierarchy of evolution equations can be then solved numerically with the arbitrary precision using one of the available ODE solvers.
\begin{widetext}
\section{Solution of evolution equations using the Laplace transform} \label{sec:sol}
\subsection{Single-excitation amplitudes} \label{sec:sol1}
 The Laplace transform $\textstyle \bar{x}(s) = \mathcal{L}_s\{x(t)\}= \int^\infty_0 \dd t \, \ee^{-s t} \, x(t)$
 turns the system of ODEs (\ref{eq:eqmot1}) governing the evolution of the single-excitation amplitudes into a set of algebraic equations as follows
 \begin{subequations} \label{eq:lapl1}
 	 \begin{gather}
 	  \left(\ii s - \tomegac\right) \bar{A}^g(s) = g \bar{A}^e(s) + f \bar{\Xi}(s), \\
 	  (\ii s - \omegaa) \bar{A}^e(s) = g \bar{A}^g(s), \\
 	  (\ii s - \omega) \, \bar{B}_\omega(s) = \xi_\omega + f \bar{A}^g(s).
 	 \end{gather}
 \end{subequations}
 The inverse Laplace transform $\textstyle y(t)=\mathcal{L}^{-1}_t\{\bar{y}(s)\}$ of the solutions of these equations gives
 	\begin{subequations} \label{eq:slns}
 	  \begin{gather}
        A^g(t) = -\ii f \frac{1}{\tR_1}\int^t_0 \dd\tau \, \Xi(\tau) \sum_{\mu=\pm} \mu (\tE^\mu_1 - \omegaa)\ee^{-\ii \tE^\mu_1 (t-\tau)}, \\
 	    A^e(t) = - \ii f g \frac{1}{\tR_1} \int^t_0 \dd\tau \, \Xi(\tau) \sum_{\mu=\pm} \mu \ee^{-\ii \tE^\mu_1 (t-\tau)}, \\
        B_\omega(t) = \xi_{\omega}\ee^{-\ii\omega t} - \ii f^2 \frac{\omega-\omegaa}{(\omega-\tE^+_1)(\omega-\tE^-_1)} \int^t_0 \dd\tau \, \Xi(\tau) \ee^{-\ii\omega (t-\tau)} + \ii f^2 \frac{1}{\tR_1} \sum_{\mu=\pm} \mu\frac{\tE^\mu_1-\omegaa}{(\omega-\tE^\mu_1)} \int^t_0 \dd\tau \, \Xi(\tau) \ee^{-\ii\tE^\mu_1 (t-\tau)}.
 	  \end{gather}
	\end{subequations} 
 Taking the long-time limit (\ref{eq:longtime}) in Eqs. (\ref{eq:slns}) and integrating over $\textstyle \tau$ one obtains
 \begin{equation} \label{eq:a_lt}
   A^{g}_\mathrm{lt} = A^{e}_\mathrm{lt} = 0, \quad B_{\omega,\mathrm{lt}} = \ee^{-\ii\omega t} \left[1 - \ii \kappa \frac{\omega - \omegaa}{\left(\omega-\mathcal{E}^+_1\right)\left(\omega-\mathcal{E}^-_1\right)}\right] \xi(\omega).
 \end{equation}
 Employing the relation $\textstyle (\omega-\mathcal{E}^+_1)(\omega-\mathcal{E}^-_1) = (\omega-\omegaa)(\omega-\tomegac)-g^2$ in the square brackets in the rhs of the expression for $\textstyle B_{\omega,\mathrm{lt}}$ one obtains
 \begin{equation}
  B_{\omega,\mathrm{lt}} = \ee^{-\ii\omega t} \frac{(\omega-\tE^+_1)^*(\omega-\tE^-_1)^*}{(\omega-\tE^+_1)(\omega-\tE^-_1)} \xi(\omega),
 \end{equation}
 which immediately leads to Eq. (\ref{eq:psi1_out}). Hereinafter, we use the subscript ``$\textstyle \mathrm{lt}$'' to indicate that the amplitude is taken in the long-time limit (\ref{eq:longtime}).
\subsection{Two-excitation amplitudes} \label{sec:sol2}
 The equations of motion (\ref{eq:eqmot2}) governing the evolution of the two-excitation amplitudes after the Laplace transform acquire the following form
 \begin{subequations} \label{eq:lapl_2}
	\begin{gather}
	 \left[\ii s - (\omega + \omega')\right] \bar{\varPhi}_{\omega, \omega'}(s) = \xi_\omega \xi_{\omega'} + \frac{f}{\sqrt{2}} \left[\bar{X}^g_\omega(s) + \bar{X}^g_{\omega'}(s)\right], \\
	 \left[\ii s - (\tomegac+\omega)\right] \bar{X}^g_\omega(s) = g \bar{X}^e_\omega(s) + \sqrt{2} f \bar{Z}^g(s) + \sqrt{2} f \bar{A}_\omega(s) * \bar{\Xi}(s), \\
	 \left[\ii s - (\omega+\omegaa)\right] \bar{X}^e_\omega(s) = g \bar{X}^g_\omega(s) + f \bar{Z}^e(s), \\
	 \left(\ii s - 2 \tomegac\right) \bar{Z}^g(s) = g \sqrt{2} \bar{Z}^e(s) + 2 f \bar{B}^g(s) * \bar{\Xi}(s), \\
	 \left[\ii s - (\tomegac+\omegaa)\right] \bar{Z}^e(s) = g \sqrt{2} \bar{Z}^g(s) + \sqrt{2} f \bar{B}^e(s) * \bar{\Xi}(s).
	\end{gather}
 \end{subequations}
 Here we have employed the convolution theorem $\textstyle \mathcal{L}_s\{y_1(t) y_2(t)\} = \bar{y}_1(s) * \bar{y}_2(s)$, where $\textstyle *$ denotes the convolution operator. Solving the above set of equations and applying the inverse Laplace transform to the obtained solutions gives the sought-for expressions for the two-excitation amplitudes in the time domain. These tedious mathematical operations were assisted by tools of \textit{Mathematica} computer algebra system \cite{trott}.
 The solution for $\textstyle \bar{\varPhi}_{\omega,\omega'}(t)$ reads
 \begin{equation} \label{eq:Phit}
  \varPhi_{\omega,\omega'}(t) = \xi_\omega \xi_{\omega'} \ee^{-\ii (\omega + \omega') t} - \ii f^2\left[\varUpsilon_{\omega,\omega'}(t) + \varUpsilon_{\omega',\omega}(t)\right],
 \end{equation}
 where $\textstyle \varUpsilon_{\omega,\omega'}(t)$ is given by
 \begin{equation} \label{eq:ups}
  \begin{split}
   \varUpsilon_{\omega,\omega'}(t) =
   & \, \sum_{\mu=\pm} \mu \frac{1}{\tR_2(\omega+\omega'-\tE^\mu_2)} \int \dd\tau \, \ee^{-\ii \tE^\mu_2(t-\tau)} \Xi(\tau) \\
   & \, \times \left[fg \frac{2(\omega + \tE^+_1 + \tE^-_1 - \tE^\mu_2)-\tE^\mu_2}{(\omega+\tE^+_1 - \tE^\mu_2)(\omega+\tE^-_1 - \tE^\mu_2)} A^e(\tau) + 2 f \frac{(\omega+\omegaa-\tE^\mu_2)(\tE^+_1+\tE^-_1-\tE^\mu_2)+g^2}{(\omega+\tE^+_1 - \tE^\mu_2)(\omega+\tE^-_1 - \tE^\mu_2)} A^g(\tau)\right] \\
   & \, - \sum_{\mu=\pm} \mu \frac{1}{\tR_1(\omega'-\tE^\mu_1)} \int \dd\tau \, \ee^{-\ii (\omega+\tE^\mu_1)(t-\tau)} \Xi(\tau) \\
   & \, \times \left[(\tE^\mu_1 - \omegaa) B_\omega(\tau) + f g \frac{(\omega+\tE^\mu_1) +   2\tE^{\bar{\mu}}_1}{(\omega+\omega'-\tE^+_2)(\omega+\omega'-\tE^-_2)} A^e(\tau) + 2 f \frac{(\tE^\mu_1-\omegaa)(\omega+\tE^{\bar{\mu}}_1)+g^2}{(\omega+\omega'-\tE^+_2)(\omega+\omega'-\tE^-_2)} A^g(\tau) \right] \\
   & \, + \frac{1}{(\omega'-\tE^+_1)(\omega'-\tE^-_1)} \int^t_0 \dd \tau \, \ee^{-\ii(\omega+\omega')(t-\tau)} \Xi(\tau) \\ & \,
   \times \left[(\omega'-\omegaa) B_\omega(\tau) + f g \frac{(\omega+\omega')+2(\omega'-\tE^+_1-\tE^-_1)}{(\omega+\omega'-\tE^+_2)(\omega+\omega'-\tE^-_2)} A^e(\tau) +
   2 f \frac{(\omega'-\omegaa)(\omega'-\tE^+_1-\tE^-_1)+g^2}{(\omega+\omega'-\tE^+_2)(\omega+\omega'-\tE^-_2)} A^g(\tau)\right].
  \end{split}
 \end{equation} 	
 Here and in what follows $\textstyle \bar{\mu}$ denotes the sign opposite to the sign of $\textstyle \mu$.
 The rest of the solutions for the two-excitation amplitudes read
 \begin{subequations} \label{eq:sltns}
  \begin{gather*}
   \begin{split} \label{eq:X_g}
    X^g_\omega(t) = & \, \ii \frac{\sqrt{2}f}{\tR_2} \sum_{\mu=\pm} \mu \int \dd\tau \, \ee^{-\ii \tE^\mu_2(t-\tau)} \Xi(\tau) \\
    & \, \times \left[fg \frac{2(\omega + \tE^+_1 + \tE^-_1 - \tE^\mu_2)-\tE^\mu_2}{(\omega+\tE^+_1 - \tE^\mu_2)(\omega+\tE^-_1 - \tE^\mu_2)} A^e(\tau) + 2 f \frac{(\omega+\omegaa-\tE^\mu_2)(\tE^+_1+\tE^-_1-\tE^\mu_2)+g^2}{(\omega+\tE^+_1 - \tE^\mu_2)(\omega+\tE^-_1 - \tE^\mu_2)} A^g(\tau)\right] \\
    & \, - \ii \frac{\sqrt{2}f}{\tR_1} \sum_{\mu=\pm} \mu \int \dd\tau \, \ee^{-\ii (\omega+\tE^\mu_1)(t-\tau)} \Xi(\tau) \\
    & \, \times \left[(\tE^\mu_1 - \omegaa) B_\omega(\tau) + f g \frac{(\omega+\tE^\mu_1) + 2\tE^{\bar{\mu}}_1}{(\omega+\omega'-\tE^+_2)(\omega+\omega'-\tE^-_2)} A^e(\tau) + 2 f \frac{(\tE^\mu_1-\omegaa)(\omega+\tE^{\bar{\mu}}_1)+g^2}{(\omega+\omega'-\tE^+_2)(\omega+\omega'-\tE^-_2)} A^g(\tau) \right],
   \end{split} \\
   \begin{split} \label{eq:X_e}
    X^e_\omega(t) = & \, \ii \frac{\sqrt{2}f}{\tR_2} \sum_{\mu=\pm} \mu \int \dd\tau \, \ee^{-\ii \tE^\mu_2(t-\tau)} \Xi(\tau) \\
    & \, \times \left[f \frac{(\omega+\omegaa-\tE^\mu_2)(2\tomegac-\tE^\mu_2) + 2g^2}{(\omega+\tE^+_1 - \tE^\mu_2)(\omega+\tE^-_1 - \tE^\mu_2)} A^e(\tau) + 2 f g \frac{\omega-\tomegac-\mu\tR_2}{(\omega+\tE^+_1 - \tE^\mu_2)(\omega+\tE^-_1 - \tE^\mu_2)} A^g(\tau)\right] \\
    & \, - \ii \frac{\sqrt{2}f}{\tR_1} \sum_{\mu=\pm} \mu \int \dd\tau \, \ee^{-\ii (\omega+\tE^\mu_1)(t-\tau)} \Xi(\tau) \\
    & \, \times \left[g B_\omega(\tau) + f \frac{(\tE^\mu_1 - \omegaa)(\omega+\tE^\mu_1-2\tomegac)+2g^2}{(\omega+\omega'-\tE^+_2)(\omega+\omega'-\tE^-_2)} A^e(\tau) + 2 f g  \frac{\omega-\tomegac+\mu \tR_1}{(\omega+\omega'-\tE^+_2)(\omega+\omega'-\tE^-_2)} A^g(\tau) \right],
   \end{split} \\
   Z^g(t) = -\ii \frac{f}{\tR_2} \sum_{\mu=\pm} \mu \int^t_0 \dd\tau \, \ee^{-\ii \tE^\mu_2 (t-\tau)} \Xi(\tau) \left[g A^e(\tau) + (\tE^\mu_2 - \tE^+_1 - \tE^-_1) A^g(\tau)\right], \label{eq:Z_g} \\
   Z^e(t) = -\ii \frac{\sqrt{2} f}{\tR_2} \sum_{\mu=\pm} \mu \int^t_0 \dd\tau \, \ee^{-\ii \tE^\mu_2 (t-\tau)} \Xi(\tau) \left[(\tE^\mu_2 - 2\tomegac) A^e(\tau) + 2 g A^g(\tau)\right].  \label{eq:Z_e} 
  \end{gather*}
 \end{subequations}

 In the long-time limit (\ref{eq:longtime}) all terms $\textstyle \propto \exp(-\ii \tE^\pm_j t)$ (where $\textstyle j=1,2$) approach zero since $\textstyle \mathrm{Im}\{\tE^\pm_j\} < 0$. This implies that all two-excitation  amplitudes except $\textstyle \varPhi_{\omega,\omega'}(t)$ turn zero at times satisfying the criterion (\ref{eq:longtime}).
 In Eq. (\ref{eq:ups}) only the last term survives in the long-time limit. Using Eqs. (\ref{eq:def_xi}) and (\ref{eq:Xi_lor}) and integrating over $\textstyle \tau$ in Eq. (\ref{eq:ups}) after a considerable amount of lengthy algebra one obtains $\textstyle \varPhi_{\omega,\omega', \mathrm{lt}} = \ee^{-\ii (\omega + \omega') t} \Phi^\mathrm{out}_{\omega,\omega'}$, with $\Phi^\mathrm{out}_{\omega,\omega'}$ given by Eq. (\ref{eq:phiout}).
\end{widetext} 
%
%

%
%
\end{document}